\documentclass[aps,prl,twocolumn,showpacs,superscriptaddress,floatfix]{revtex4}

\usepackage{graphicx}

\begin{document}

\title{Visualizing the Spin of Individual Molecules}
\author{C.\ Iacovita}
\author{ M.\ V.\ Rastei}
\author{B.\ W.\ Heinrich}
\affiliation{Institut de Physique et Chimie des Mat\'{e}riaux de Strasbourg$\text{,}$ UMR 7504$\text{,}$ Universit\'{e} Louis Pasteur, F-67034 Strasbourg, France}
\author{T.\ Brumme}
\author{J.\ Kortus}
\affiliation{Institut f\"{u}r Theoretische Physik, TU Bergakademie Freiberg,  D-09599 Freiberg, Germany}
\author{L.\ Limot}
\email{limot@ipcms.u-strasbg.fr}
\author{J.\ P.\ Bucher}
\affiliation{Institut de Physique et Chimie des Mat\'{e}riaux de Strasbourg$\text{,}$ UMR 7504$\text{,}$ Universit\'{e} Louis Pasteur, F-67034 Strasbourg, France}

\date{\today}

\begin{abstract}
Low-temperature spin-polarized scanning tunneling microscopy is employed to study spin transport across single Cobalt-Phathalocyanine molecules adsorbed on well characterized magnetic nanoleads. A spin-polarized electronic resonance is identified over the center of the molecule and exploited to spatially resolve stationary spin states. These states reflect two molecular spin orientations and, as established by density functional calculations, originate from a ferromagnetic molecule-lead superexchange interaction mediated by the organic ligands.  
\end{abstract}


\maketitle 
Conceptually new device structures accounting for sizable quantum effects will be needed if the downscaling of electronic and magnetic devices were to continue. One of these new concepts is the marriage of molecular electronics and spintronics, where functional molecules become active device components within a circuitry where information is carried by spins \cite{roc_05,bog_08}. Progress toward this tantalizing goal rely on our understanding of spin transport and magnetism in reduced dimensions, where fundamental playgrounds extend to the extreme limit of single atoms and molecules. Studies include spin transport across a well chosen molecule sandwiched between two magnetic leads \cite{tsu_99,pet_04}, or even atomic-size constrictions formed by bringing two leads into contact \cite{cho_05,sok_07}. One of the limiting drawbacks is the variability of the resulting conductance, which comes from the incomplete knowledge we have of the molecule-lead interface. For instance, only a few experimental studies have focused on the interaction between a molecular spin and a magnetic substrate \cite{sch_05,wen_07}. A better understanding would enable us to target the chemical engineering needed for building the desired spintronic functionalities into a molecule.\\
\- In the past years, it became possible with spin-polarized (SP) scanning tunneling microscopy and spectroscopy (STM and STS) to directly observe the interplay between magnetism and surface structure with atomic resolution. SP-STM can also serve as a model tunneling magnetoresistance device since the junction includes two well defined magnetic leads -- the tip and the sample -- separated by a vacuum barrier. A link can then be accurately established between spin transport and density of states. Recently, through SP-STM it was possible to evidence how the magnetization switching of a nanocluster is influenced by the spatial location of the spin injection \cite{kra_07}. Another example can be found in SP-STM of single atoms \cite{yay_07}. When the SP current flows across a magnetic atom adsorbed on a magnetic surface rather than directly into the surface, the tunneling spin transport is significantly affected, and some control can be exerted through the choice of the atom.\\
\begin{figure}[b]
  \includegraphics[width=0.35\textwidth,clip]{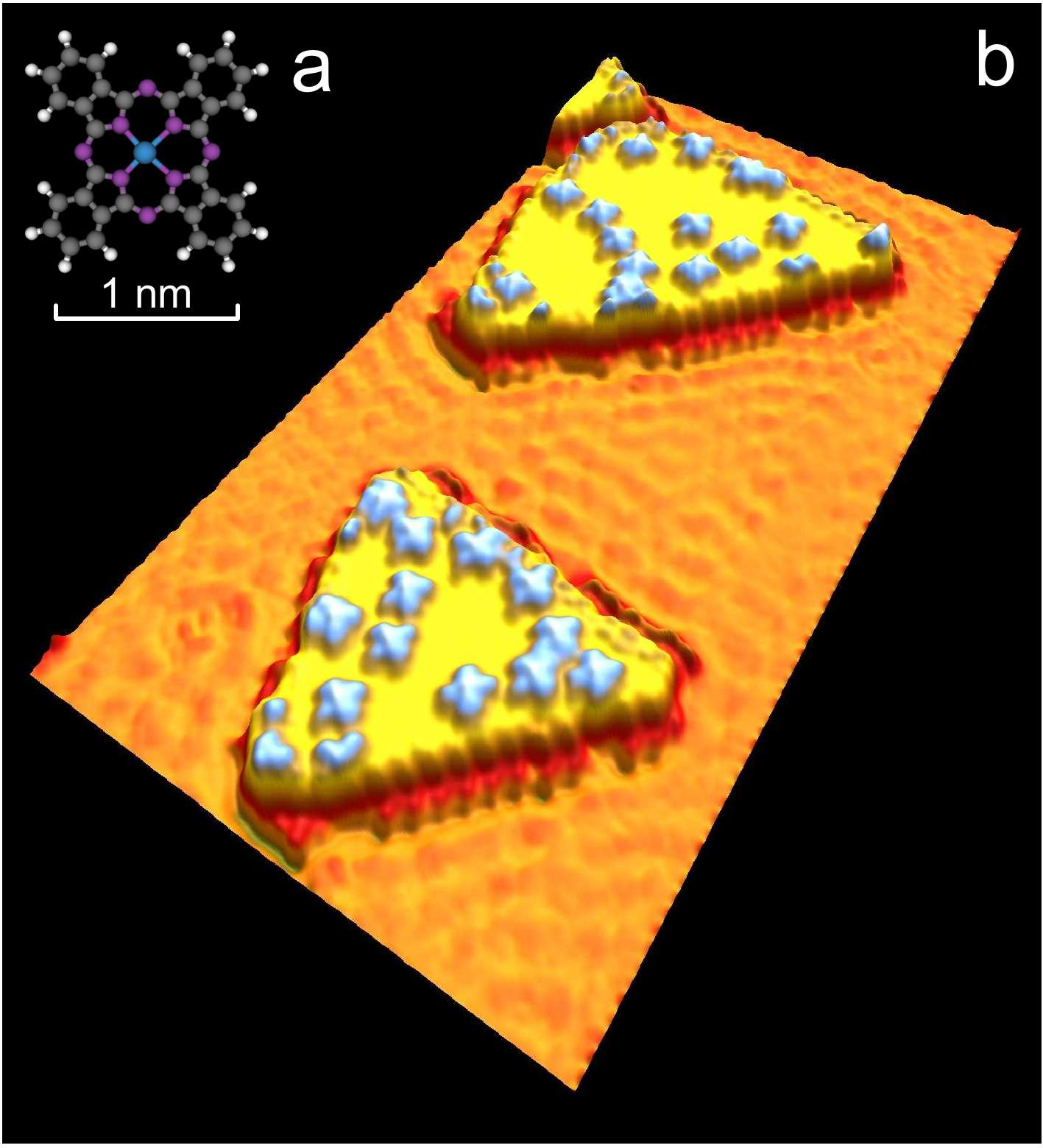}
  \caption{CoPc molecules adsorbed on cobalt nanoislands grown on Cu(111). a) Structure model for CoPc (hydrogen: white, carbon: gray, nitrogen: violet, cobalt: blue). b) Pseudo three-dimensional STM image ($40\times 20$ nm$^2$, $0.1$ V, $0.5$ nA). The spatial oscillations on the Cu(111) surface are due to the scattering of the Shockley surface-state electrons.}
\label{fig1}
\end{figure}
Here we show how tunneling spin transport can be modified by ``dressing'' atoms with organic ligands. We combine low-temperature SP-STM and model calculations to study a model system consisting of individual Cobalt-Phthalocyanine (CoPc, Fig.~\ref{fig1}a) molecules adsorbed on a magnetic substrate. The interaction between the molecule and the substrate results in two molecular spin orientations clearly discernible in the differential conductance of the SP tunnel junction ($dI/dV$). Our results provide a direct visualization of stationary spin states, that differs from the molecular Kondo state where spin flips occur in time \cite{wah_05,zha_05,ian_06,gao_07,fu_07}. As demonstrated here this is a consequence of the ferromagnetic coupling between the molecular spin and the magnetic substrate.\\
\begin{figure}[t]
\includegraphics[width=0.4\textwidth,clip=]{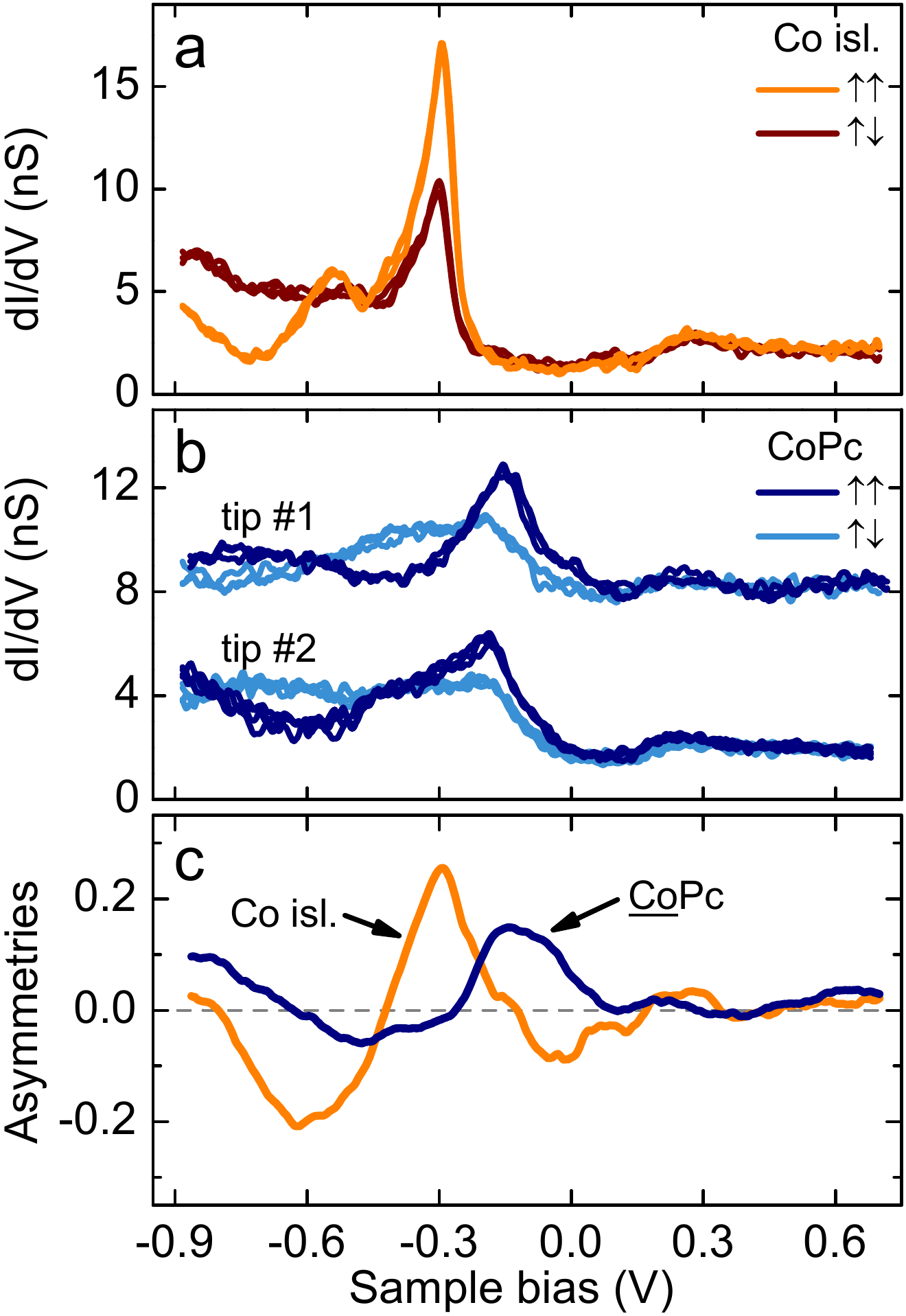}
\caption{Spin-polarized tunneling conductance through the nanoislands and CoPc. a) Typical spin-polarized $dI/dV$ over two cobalt nanoislands of opposite magnetization (orange: $\uparrow\uparrow$, brown: $\uparrow\downarrow$). Feedback loop opened at $0.6$ V and $0.5$ nA. b) Differential conductance ($dI/dV$) over the centre of single CoPc molecules adsorbed on cobalt nanoislands of opposite magnetization (dark blue: $\uparrow\uparrow$, light blue: $\uparrow\downarrow$). Two sets of spectra acquired with distinct tips (noted 1 and 2) are presented. The spectra acquired with tip 1 are displaced upward by $6$~nS for clarity. Feedback loop opened at $0.6$~V and $0.5$~nA. c) Asymmetries arising from opposite magnetizations: CoPc (dark blue) and Co nanoislands (orange). The asymmetries are an average of all the recorded asymmetries obtained with different tips.}
\label{fig2}
\end{figure}  
The measurements were performed in a STM operating below $10^{-10}$~mbar at a temperature of 4.6 K. The magnetic substrate was obtained by evaporating cobalt onto a pristine Cu(111) single crystal held at room temperature, leading, at submonolayer coverage, to the formation of triangular-like nanoislands two-atomic layers high (Fig.~\ref{fig1}b). Cobalt islands were chosen since they are a well studied system \cite{pie_04,ras_07}, with, at low temperature, a magnetization perpendicular to the surface. In order to sense the spin-dependent contribution through the differential conductance, Co-coated metal tips with an out-of-plane spin-sensitivity were employed.
The $dI/dV$ versus sample bias ($V$) was recorded by superimposing a sinusoidal modulation to the junction bias of amplitude $5$~mV~rms at a frequency of $7$~kHz, and detecting the first-harmonic of the current through a lock-in amplifier. Low-temperature constant-current images showed that CoPc adsorbs preferentially on cobalt, either on top of the nanoislands or along their step edges (Fig.~\ref{fig1}b). Typical island coverage was roughly $10$ molecules for an area of $(10$~nm$)^{2}$. On the islands, CoPc exhibits a four-lobe pattern $1.4$~nm wide consistent with the ideal structure of the molecule.\\  
\- The spectra of the differential conductance ($dI/dV$) versus sample bias in the center of the islands in the absence of CoPc are marked by a dominant SP resonance falling at $-0.28$~V below the Fermi energy (Fig.~\ref{fig2}a), which is believed to arise from the hybridization of $s-p$ states with the minority $d_{3z^2-r^2}$ band \cite{ras_07}. Its amplitude is sensitive to the spin polarization of the Co islands, so that islands with opposite magnetization can be discerned through single spectroscopy or through a contrast in spectroscopic maps (see Fig.~\ref{fig3}b where $V=-0.29$~V). In the following, islands presenting a strong $dI/dV$ signal at $-0.28$~V are arbitrarily designated as parallel (noted $\uparrow\uparrow$), while the remaining as antiparallel ($\uparrow\downarrow$). All the spectra presented were recorded over islands of \textit{hcp} stacking with lateral dimensions larger than $12$~nm. This ensures that the spectroscopic contrast is of SP nature, since structural effects as stacking \cite{pie_04} or size-dependent relaxations \cite{ras_07} are eliminated.\\
\- Spin-polarized spectra of CoPc were acquired by positioning the tip over the Co of the molecule (Fig.~\ref{fig2}b). A broad resonance centered at $-0.19$~V is detected for all CoPc molecules residing on $\uparrow\uparrow$ islands, whereas a resonance of weaker amplitude is found for molecules on $\uparrow\downarrow$ islands, indicating that the cobalt atom is spin-polarized. Here after, CoPc molecules are noted $\uparrow\uparrow$ if residing on parallel islands, and $\uparrow\downarrow$ if residing on antiparallel islands. The spin polarization of a single CoPc is therefore linked to the one of the island, which then indicates that its direction may be switched by inverting the island magnetization. This is particularly appealing in view of spin-dependent molecular electronics as suggested recently by a study on ultrathin films of metalloporphyrin \cite{wen_07}.  While the resonance is fairly reproducible, the structure at lower biases was found to be tip dependent (Fig.~\ref{fig2}b). To get a clearer picture, we have plotted the asymmetries of both CoPc (blue line on Fig.~\ref{fig2}c) and the cobalt islands (orange line) averaged over 13 different tips. The asymmetry is defined as $(\uparrow\uparrow-\uparrow\downarrow)/(\uparrow\uparrow+\uparrow\downarrow)$, 
the arrows referring to a $dI/dV$ acquired on $\uparrow\uparrow$ and $\uparrow\downarrow$ CoPc molecules or islands, respectively. An oscillatory behavior is observed with a sign reversal occurring at $-0.26$ and $-0.64$~V for CoPc, and at $-0.12$, $-0.42$ and $-0.80$~V for the islands.\\
\- A visual rendering of the $dI/dV$ spectra is presented on Fig.~\ref{fig3}, where the pixel intensity on each image corresponds to the differential conductance at a given sample bias. These images, or $dI/dV$ maps, show isolated CoPc molecules on two islands of opposite contrast. A marked island contrast (areas without CoPc) is visible in Fig.~\ref{fig3}a since the bias is then $-0.29$~V which corresponds to a high magnetic asymmetry (see Fig.~\ref{fig2}c). The right island is $\uparrow\uparrow$, whereas the left one is $\uparrow\downarrow$. A similar contrast is observed in the absence of CoPc (Fig.~\ref{fig3}b). On the contrary, CoPc molecules have similar intensities on both islands, as expected by the molecular asymmetry which is close to zero at this bias. A contrast between $\uparrow\uparrow$ and $\uparrow\downarrow$ molecules becomes visible when moving upward or downward in energy, as one may expect from the behavior of the asymmetry depicted in Fig.~\ref{fig2}c. While the islands have nearly same intensity at $-0.16$ V (Fig.~\ref{fig3}c), the CoPc asymmetry is now maximal. The molecules appear as round dots, with a higher intensity for CoPc molecules residing on the $\uparrow\uparrow$ island compared to CoPc on the $\uparrow\downarrow$ island (Fig.~\ref{fig3}d). A contrast is also detected at $-0.32$ V (Fig.~\ref{fig3}e), but is now reversed compared to the previous map. The benzene ring of the organic ligand (Fig.~\ref{fig3}f) appears here as a dim extension of a bright area centered at the Co atom of full width at half maximum of $0.5$~nm. We observed that the SP signal, when present, is spatially limited to this area (see Fig.~S1 in the auxiliary information).\\
\begin{figure}
\includegraphics[width=0.45\textwidth,clip=]{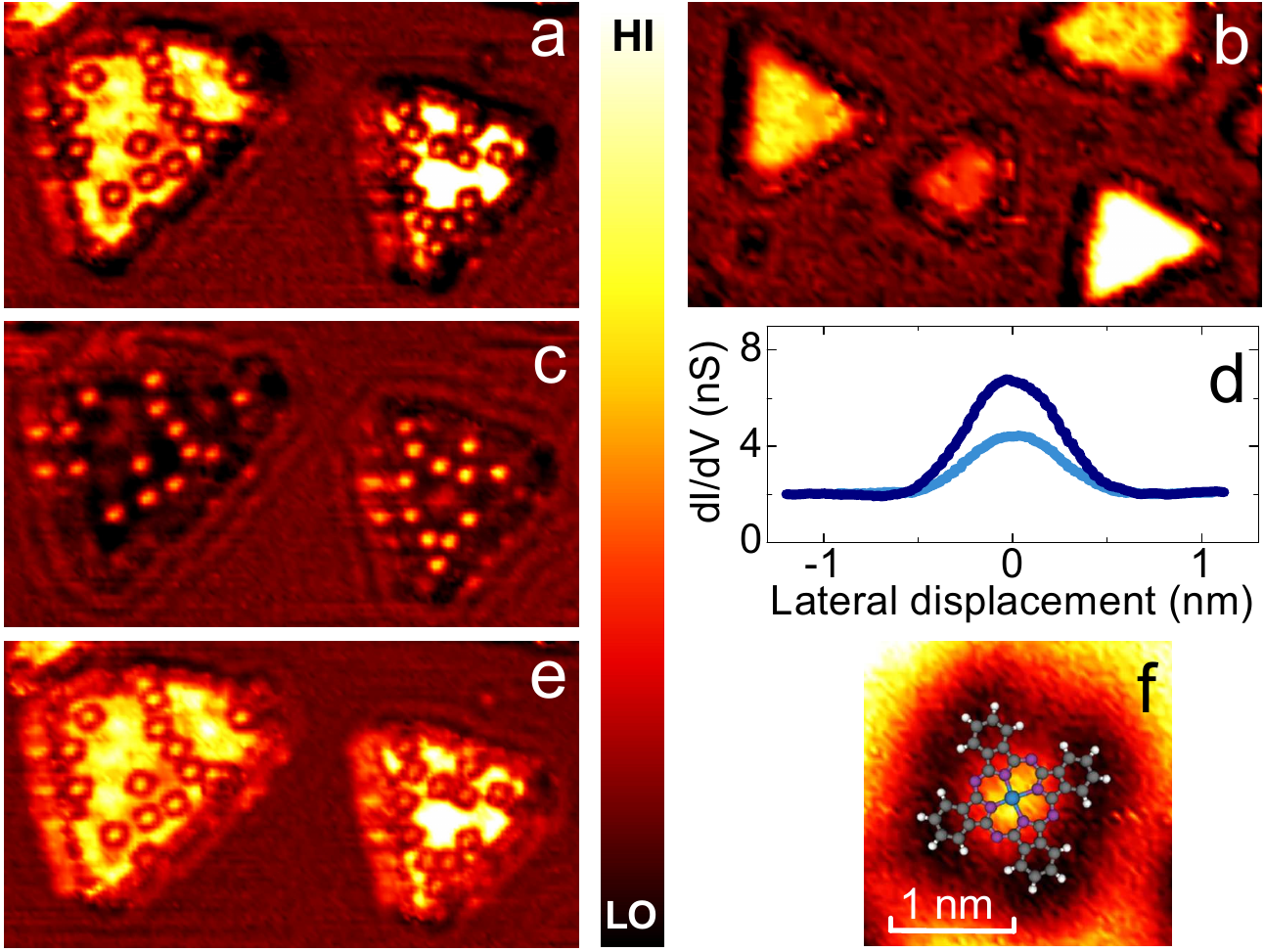}
\caption{Spin-polarized tunneling conductance maps for CoPc. Maps a), c), e) were taken at $-0.29$, $-0.16$, and $-0.32$~V respectively (image size: $40\times 20$ nm$^2$). Each map is normalised to span the same colour palette range. Feedback loop opened at $0.6$~V and $0.5$~nA. b) Spin-polarized map of Co nanoislands at $-0.29$ V -- as in a) -- in the absence of CoPc molecules ($45\times 25$ nm$^2$). d) Profiles of the differential conductance along one axis of CoPc extracted from c) (dark blue: $\uparrow\uparrow$, light blue: $\uparrow\downarrow$). f) Map of a $\uparrow\downarrow$ molecule of e) with model molecular structure superimposed ($2.6\times 2.6$ nm$^2$).}
\label{fig3}
\end{figure}
In order to support the experimental findings of a twofold spin orientation for CoPc and gain some insight on the magnetic coupling with the island, we carried out first-principles calculations based on density functional theory by means of the \textsc{PWscf} package \cite{pwscf}. A detailed presentation of the method is provided as auxiliary information. The magnetic surface is mimicked by using a slab model consisting of three Cu layers with two Co layers on top and about $2.2$~nm vacuum to the next periodic repeated layer. The CoPc was placed above the Co-layers resulting in a 302 atom model. During the geometry optimization only the Cu atoms where held fixed at their positions according to bulk values, whereas CoPc and the two Co-layers where allowed to fully relax. Different starting geometries with the cobalt atom of CoPc in a top, bridge and hollow position have been checked. The lowest energy from the geometry optimization has been obtained for the bridge position. The calculated distance between the surface atoms and the cobalt atom of CoPc is about $0.25$~nm, close to the Co-Co distance in the surface layers, with only a small distortion of the surface occurring (Fig.~S2).\\
\- The spin-polarized partial density of states (PDOS) gives information on possible hybridizations among atoms of CoPc and with the surface atoms (Fig.~\ref{fig4}a). The broad DOS between $-4$~eV and the Fermi level ($E_F$) is dominated by the hybridization of the $d_{\pi}$ orbital ($d_{xz}, d_{yz}$) with N $p_z$-states (explicitly shown by a green solid line on Fig.~\ref{fig4}b). The remaining states are realized by $d_{x^2-y^2}$- and $d_{xy}$-states (dark red line on Fig.~\ref{fig4}b). The prominent minority peak at $-2$~eV corresponds to $d_{xy}$. A bonding combination of $p$-N and $d_{x^2-y^2}$ falls at $-6$~eV for majority states (Fig.~\ref{fig4}a). A small hybridization between cobalt surface atoms and nitrogen/carbon atoms of CoPc providing the chemical bonding to the surface is found at binding energies between $-4$ to $-5$~eV. Due to this interaction the CoPc is no longer planar (Fig.~S2). The $d_{z^2}$ state shows strong hybridization with $p$-N which distributes it over several occupied states (dark blue line on Fig.~\ref{fig4}b). In contrast to a free CoPc molecule, where the spin down lowest unoccupied molecular orbital (LUMO) can be identified as $d_{z^2}$ \cite{zha_05}, it becomes occupied here. This results in a reduced magnetic moment of about $0.7$~$\mu_B$ for CoPc adsorbed on the cobalt island. In order to obtain a more detailed insight into the chemical bonding one can analyze the wavefunctions or charge densities within an energy window, which reveal that there is also direct overlap between the surface Co surface atoms and the Co atom of CoPc, as shown in Fig.~S3. The PDOS shows in particular that the resonance at $-0.19$~V detected by SP-STS over the cobalt atom of CoPc likely originates from a mixture of minority $d_{z^2}$ and $d_{\pi}$ states (Fig.~\ref{fig4}b), the contribution of $d_{x^2-y^2}$ and $d_{xy}$- states being negligible near $E_F$. This conclusion should also hold in the vacuum region where the tip is, with however a possible change in the relative amplitudes of the $d_{z^2}$ contribution with respect to the $d_{\pi}$ one, since states with a small in-plane $\overrightarrow{k}$-component are the states that show the slowest decay into vacuum, $i.\ e.$ the states to which STM is the most sensitive.\\
\begin{figure}
\includegraphics[width=0.45\textwidth,clip=]{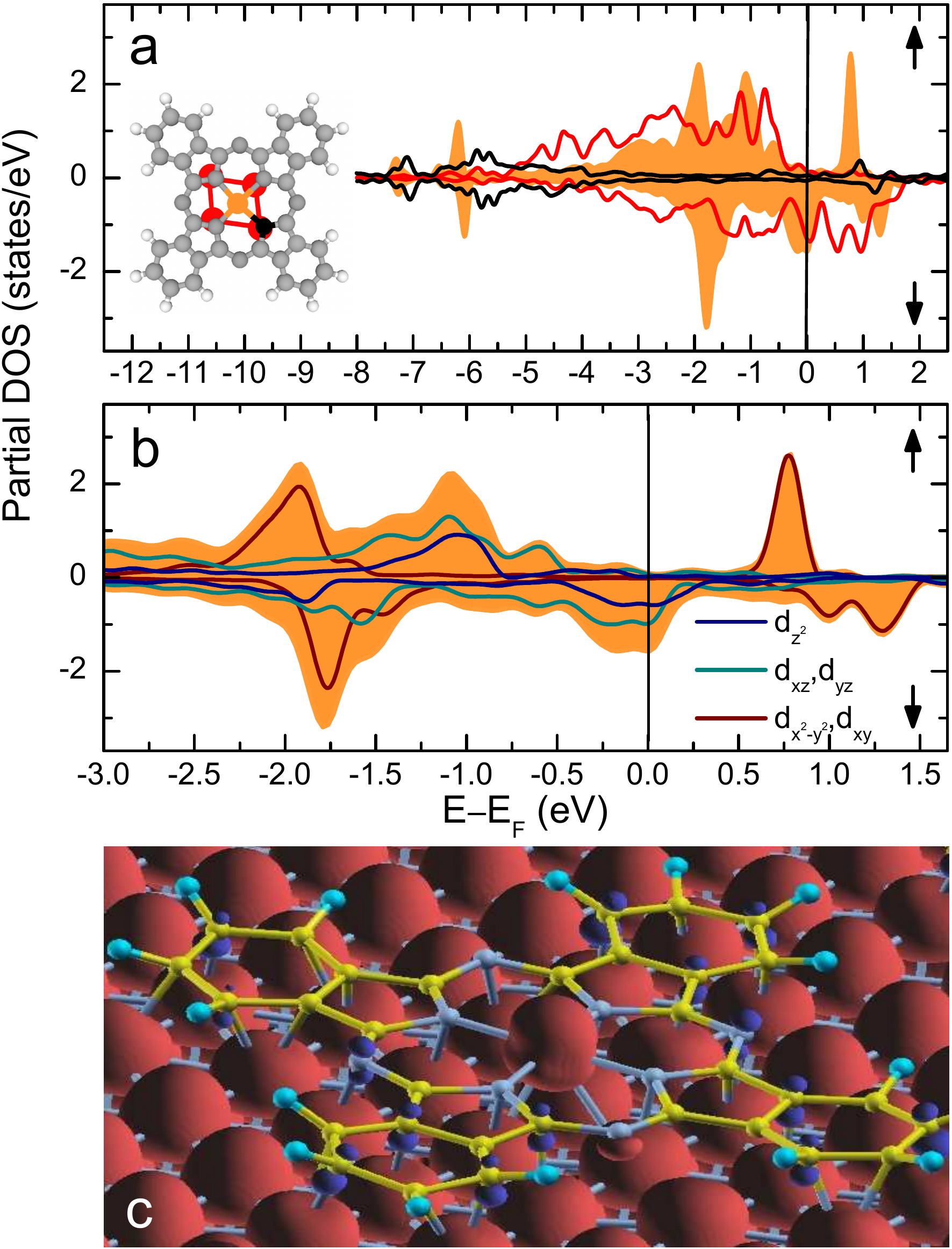}
\caption{Electronic and magnetic structure of CoPc on the  cobalt nanoislands. a) Spin-resolved density of states (DOS) of the Co atom of CoPc (orange line), of a nitrogen atom (black line), and of a cobalt atom of the surface with the Co-N-CoPc angle close to 89.5$^{\circ}$ (red line). Inset: Schematic adsorption geometry of CoPc. b) Spin-resolved DOS of the Co atom of CoPc and contribution from the $d$-states involved. c) Isosurface plot of the magnetization density (red colour corresponds to spin up and blue spin down). The CoPc is coupled ferromagnetically to the surface. There is a spin polarization on some carbon and nitrogen atoms.}
\label{fig4}
\end{figure}
The magnetization density (Fig.~\ref{fig4}c) shows clear ferromagnetic coupling between CoPc and the magnetic surface. While the largest magnetization density is found close to the cobalt atoms, as expected, there is also some very small magnetization density at some nitrogen and carbon atoms ($<0.05$~$\mu_B$), which might explain why a SP signal could not be evidenced over these atoms. Some of the outer ligand atoms of CoPc show antiparallel alignment of the magnetization, which is characteristic for spin polarization through these atoms. The bridge position is driven by the interaction of the four N atoms around the central cobalt atom of CoPc (inset of Fig.~\ref{fig4}a and Fig.~S2). The geometry optimization reveals that one of the N atoms is on top of a surface atom. The remaining three N atoms are in less favorable positions due to the misfit of the molecular geometry and the cobalt surface below. The N atom on top of a surface atom should show the largest wavefuction overlap. The angle between the Co surface atom, the N atom on top and the cobalt atom from CoPc is 89.5$^{\circ}$, which would favor ferromagnetic exchange according to the Goodenough-Kanamori rules for superexchange \cite{good_63}. This indirect exchange coupling mechanism is in accordance with the conclusions drawn for metalloporphyrins \cite{wen_07}, here, however, the direct Co-CoPc distance is comparable to the average distance within the cobalt surface, so that direct exchange may contribute as well but in a weaker extent.\\
\- With the ultimate lateral resolution of SP-STM we have shown that the spin state of a single molecule can be directly visualized by means of spin-polarized tunnel electrons. This is possible because a spin-polarized resonance built on $d$ molecular orbitals is found close to the Fermi level. Additionally the orientation of the molecular spin is fixed in time owing to a ferromagnetic coupling with the magnetic substrate. The organic ligands play an important role in both aspects, suggesting that control over spin transport may be achieved through coordination chemistry. Our results also open the exciting perspective of observing changes in single-molecule magnetism induced by external stimuli such as light or a magnetic field.\\
\- This work was supported by the European Union Network of Excellence MAGMANet (FP6-515767-2). We would like to thank the ZIH Dresden for providing computational resources and assistance.

\end{document}